\documentclass{article}

\usepackage{PRIMEarxiv}

\usepackage[utf8]{inputenc} 
\usepackage[T1]{fontenc}    
\usepackage{hyperref}       
\usepackage{url}            
\usepackage{booktabs}       
\usepackage{amsfonts}       
\usepackage{nicefrac}       
\usepackage{microtype}      
\usepackage{lipsum}
\usepackage{fancyhdr}       
\usepackage{graphicx}       
\graphicspath{{media/}}     
\usepackage{float} 
\title{CBIM: object-level cloud collaboration platform for supporting across-domain asynchronous design}

\author{
Zijian Wang, Boyuan Ouyang, Rafael Sacks\\
  VClab \\
  Technion – Israel Institute of Technology\\
  Haifa\\
  \texttt{\{zijian.wang, boyuan-terry\}@campus.technion.ac.il} \\
  \texttt{cvsacks@technion.ac.il} \\
}

\begin{document}
\maketitle

\begin{abstract}
The conventional approach of designing BIM projects requires packaging of building information as files to exchange designs. This study develops a series of components to implement a previously established Cloud BIM (CBIM) platform that facilitates fileless cloud collaboration across BIM disciplines.  A CBIM connector was developed to synchronize design changes from one discipline client to a CBIM server. The server processes the modifications and propagates relevant reference geometries to affected disciplines to assist their designs. The success of the case study demonstrates a fileless approach for multidisciplinary BIM collaboration and validates the practicality and capabilities of the CBIM paradigm.
\end{abstract}


\section{Introduction}
The Architecture, Engineering, and Construction (AEC) industry is characterized by its fragmented nature in the sense that numerous stakeholders with diverse specialties are involved throughout the lifecycle of a building project. This status quo necessitates close collaboration between different design entities, each with its methods of working and preferences for BIM authoring tools, during the design phase of projects to deliver a holistic building design. The communication process often involves extensive exchange of design models, and the collaboration is built around evolving the federated discipline models while maintaining their consistency throughout the process until the models mature for construction.

As is evident to any experienced user, objects constituting the federated building models are implicitly related to one another in many ways since they are partial representations of the same physical building. However, the practice of sharing information via vendor-neutral model files – the most widely accepted solution for BIM design exchange to date – is limited because models are not associated with meaningful semantics, thus losing the ability to incorporate intelligent object interactions. The file-based nature introduced by the model exchange practice also complicates the workflow, especially when multiple design alternatives and different versions arise. The resulting sequential workflow is characterized by inefficiency and was identified as a significant cause of design clashes and rework \cite{oraee2019collaboration, sierra2019bim}.

In response to the outlined challenges, a novel CBIM paradigm was proposed \cite{sacks2022toward} which primarily involves the use of knowledge graph representations of BIM models and semantic enrichment techniques to connect federated discipline models, once isolated and standalone, into an interrelated building information web with explicit design intent semantics. The resulting CBIM metagraph, ideally, would facilitate the deployment of intelligent software methods that operate along chains of enriched association semantics to provide users with intelligent functions capable of automating design collaboration processes.
The previous study \cite{sacks2022toward} laid down the theoretical basis and provided a projected implementation framework for CBIM. By far, it remains a conceptual idea with no solid validation or prototype implementation. This paper, therefore, exploits the CBIM concept and attempts to develop a prototypical application of CBIM in supporting fileless, asynchronous BIM design collaboration under a multidisciplinary setting. 
 
\section{Literature Review}

\subsection{File-based and object-based BIM collaboration}
Cloud-based common data environments (CDE), such as Trimble Connect \cite{trimbleconn} and BIM Collaborate \cite{bimcollab}, are commercial BIM collaboration solutions targeted for the AEC sector. Despite being developed by different vendors, these tools share many features in common. At the core, they provide centralized cloud storage space for 2D, 3D, point cloud and geospatial files that every authorized user can access. These platforms usually support a number of open data formats for 3D models like STEP and IFC, as well as a few proprietary data formats native to software under each company’s ecosystem. In essence, these centralized data repositories with built-in model viewers are designed to regulate BIM exchange processes by hosting multidisciplinary building models and related documents, allowing users to access the latest information from the “single source of truth” and view/navigate all building information conveyed in the 3D models. 

Notwithstanding the capability of cloud-based CDEs in mitigating the complex and disordered characteristics inherent in conventional AEC design exchanges, the resolution of coordination is confined to a file level. As a result, CDE platforms can only provide auxiliary collaboration functions such as file-level version control and model-level clash detection using solid geometry. Coordinating designs and resolving conflicts remain largely manual processes, making human intervention indispensable for coordinating federated building models.

On the other hand, ArchiCAD \cite{archicad} enables multidisciplinary design by offering an object-based, all-in-one solution that integrates collaboration functionalities into the authoring tool. The “single model approach” of ArchiCAD provides multidisciplinary users with a master model that can be hosted on a central server for all project participants, and everybody is involved in the same model from an early stage of the design. A significant benefit is avoiding repetitive designs or overlapped modeling of the same physical objects by more than one project team. In other words, many design conflicts can be avoided since every object and object part is only modeled once by its responsible design team. The object-based approach of ArchiCAD also offers object-level coordination functionalities to users from disparate disciplines. For example, to avoid design conflict when co-authoring a design project in parallel, users can freeze one or a set of model elements before modifying them by reserving them from other users and claim temporary ownership. Apparently, the above-mentioned benefits of ArchiCAD rely heavily on all major design teams adopting the same BIM tool and a tight collaboration among the teams to negotiate designs at a high resolution. Efforts are still needed to explore the technology to realize object-level coordination across models authored by different BIM software from different software vendors.

\subsection{Compilation and representation of BIM graphs}
Over the past decade, standardizing object-based graph representations of BIM models has been a main research goal of the Linked Building Data (LBD) community. To achieve this goal, they conceptualized domain knowledge into a set of ontologies in a machine-readable format \cite{pauwels2022knowledge}. Examples of these ontologies are ifcOWL – an ontological alternative of the IFC schema \cite{pauwels2016express}, building topology ontology (BOT) – which describes the core topological structure of a building \cite{rasmussen2021bot}, Brick – which presents physical, logical and virtual assets in buildings and their relationships \cite{balaji2016brick}, etc. These building representations devised by the LBD community follow the semantic web and linked data technologies, i.e. using the subject-predicate-object approach to describe all sorts of building information as Resource Description Framework (RDF) graphs \cite{pauwels2016express}. As a result, BIM graphs that follow the LBD approach can leverage a set of tools from the semantic web for graph construction, representation, visualization, and query.

Researchers gradually realized that compiling all building information into graphs is not practical, since not all data were appropriate for graph representation. The complex constructive solid geometry (CSG) of building elements is an example of such \cite{pauwels2022knowledge}. A more convenient approach would be to maintain these data in their original formats and associate them with the core building semantic graph for efficient retrieval. Then, researchers adopted the concept of CDEs that can be used as hubs to host heterogeneous building information from different domains. A study in this field is the graph-based core-extension data framework presented by \cite{ouyang2022semantic}. The framework splits the data repository into two layers: 1) a core graph layer to represent essential building semantics, including building objects and their relationships, according to the standard ontologies; 2) an extension layer for supplementary data in various formats that can be linked virtually to objects from the core graph layer. Another similar work, LBDserver, stores heterogeneous building data in a federated web-based CDEs by following the industry standard Information Container for linked Document Delivery (ICDD) \cite{malcolm2021lbd}. These studies provide uniform semantics for representing building model information and an efficient structure for data storage. 

The research mentioned above primarily focuses on standardizing the graphs and exploring frameworks for data representation, while practical tools are needed to compile raw building information into graphs. These tools were named parser, converter, compiler, and connector in different scenarios. One pioneer work, the IFCtoRDF parser, compiles IFC models as RDF graphs while keeping all original information by following the IFC schema \cite{ifctordf}. However, the output graph from IFCtoRDF is redundant and too complex for any practical use since it integrates every piece of data from the original model \cite{pauwels2017simplebim}. To tackle this problem, \cite{rasmussen2021bot} devised the Building Topology Ontology (BOT) and other extension ontologies to convert IFC models into modular LBD graph. The resulting building representation is more concise and lightweight than the full IFC graph, but is considered incomplete due to the deliberate trimming of data. \cite{ouyang2022semantic} presented an approach to construct a graph-based CDE by utilizing the IFCtoLBD converter to compile the core graph layer and extract object geometries from the original IFC models to form the extension layer. As information is lost during the process of saving IFC models \cite{venugopal2015ontology}, the output graphs from all current IFC-based parsers face the risk of information incompleteness. Moreover, \cite{djuedja2021integrated} designed the MINDOC as a Revit add-in to compile LBD graphs from Revit directly, while it is not open for public use. 

\begin{figure*}[ht] 
	\centering
	\includegraphics[width=1\textwidth]{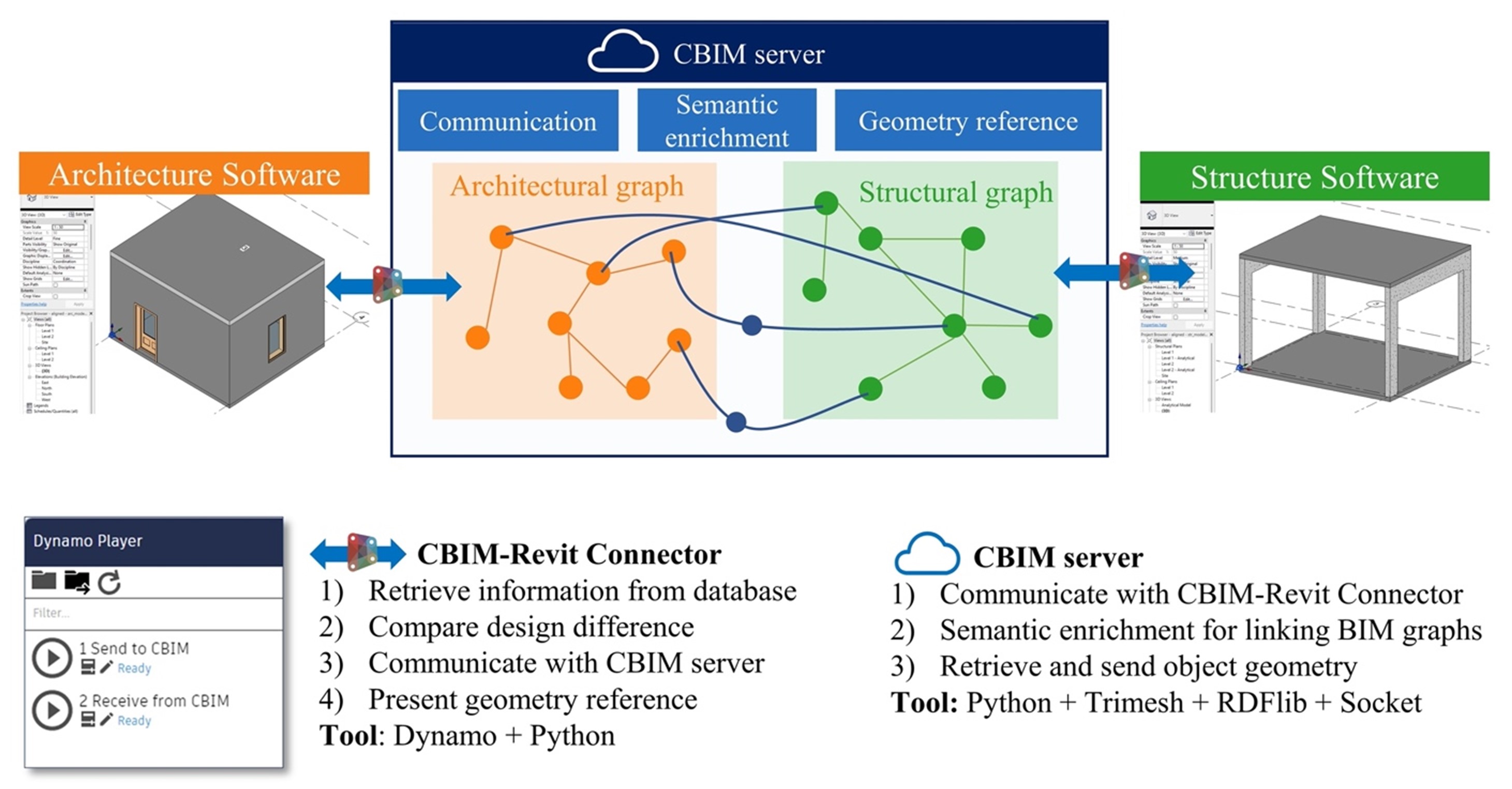}
	\caption{The CBIM platform, consisting of a CBIM-Revit connector and a CBIM server.}
	\label{fig:fig1}
\end{figure*}

\subsection{Research gaps and objectives}
While many studies have explored the theory of object-level cloud collaboration, there is no mechanism for assisting design, nor there are practical experiments and prototypes. In addition, there are no tools that can generate graph-based CDEs from BIM editors and connect to the cloud server.

This study aims to construct a series of tools to implement the CBIM collaboration platform for achieving the function of assisting across-domain asynchronous design at the object level. Specifically, this study explores the mechanism that can assist design across disciplines and develops a connector for constructing graph-based CDEs and communicating between software and the server.

\section{CBIM Platform}

In this chapter, we elucidate the main system components of the CBIM platform and explain the details of their construction. As detailed in Figure~\ref{fig:fig1}, the platform consists of a CBIM server that integrates all the software modules for the CBIM operations, as well as a CBIM-Revit Connector that maintains communication between the server and local BIM clients. This study adopts Revit as the local BIM editor for illustration purposes since the Dynamo API it incorporates offers a handy programming interface to manipulate and extract information from the building models. One should be reminded that the principles and the mechanisms devised in this study are universal and not constrained by any specific BIM editor.

\subsection{Revit-CBIM connector}

The Revit-CBIM connector has four main functions: 1) Accessing and retrieving building information inside the Revit database. 2) Comparing two versions of design models and detecting changes. 3) Sending and receiving information to and from the CBIM server. 4) Rendering reference geometries within the client interface. The connector is implemented using Dynamo functions and Python codes.

\begin{figure*}[hbt!]
	\centering
  	\includegraphics[width=\textwidth]{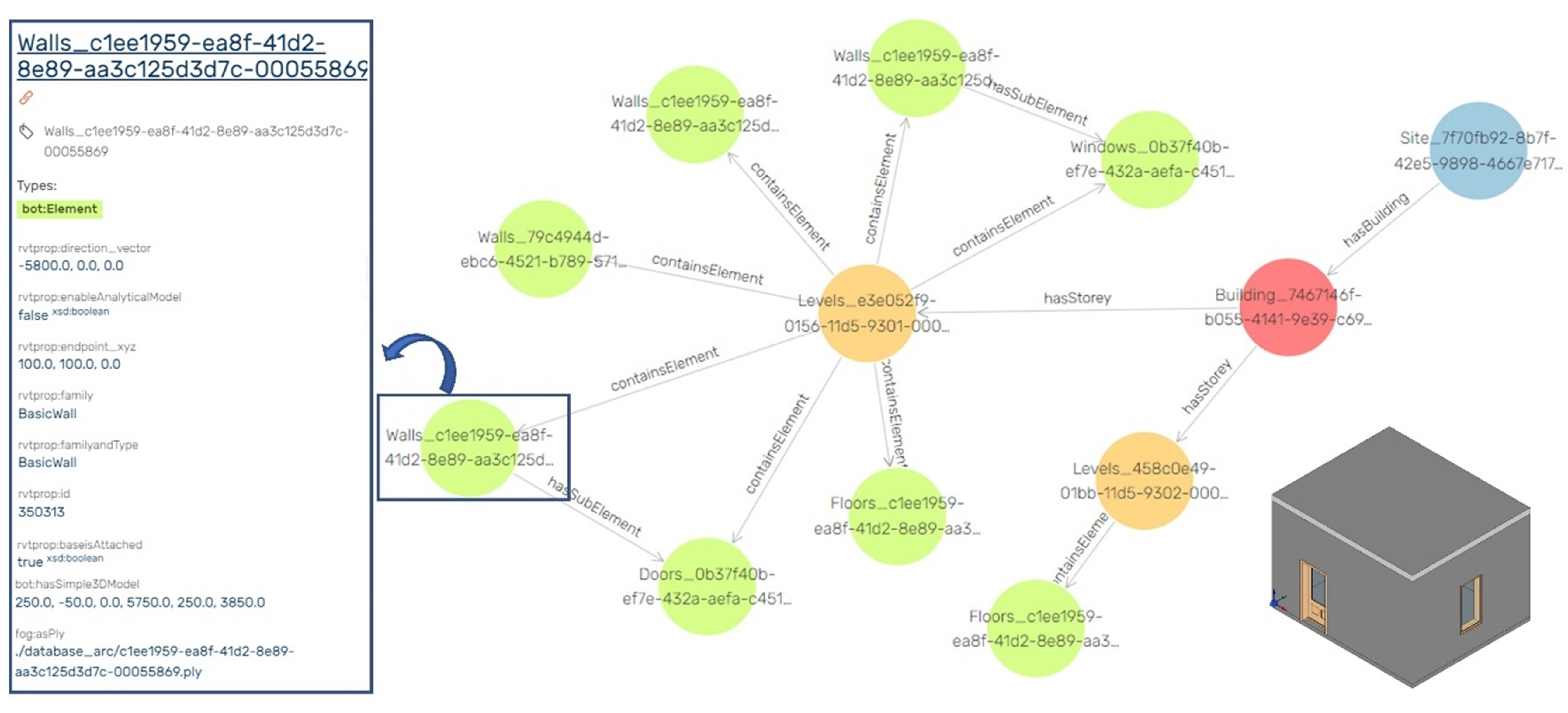}
  	\caption{The compiled graph from Revit-CBIM connector.}
  	\label{fig:fig2}
\end{figure*}

The core function of the connector is to represent information from building projects as a graph-based CDE by adopting LBD open ontologies and the core-extension structure \cite{ouyang2022semantic}. In the core graph layer, nodes represent BIM objects and links are relations among BIM objects following the BOT ontology \cite{rasmussen2021bot}. The extension layers contain the collection of exact geometries. The geometry file is named after the GUID of the corresponding object and linked to the object node following the FOG ontology \cite{bonduel2019including}. 
In practice, object geometries are saved as PLY files and their addresses are represented as an attribute in the corresponding node. Whenever geometry information is needed, users can query the core graph and retrieve the geometry file by accessing the file address. In addition, the connector maintains original attribute names from Revit. An example of the core graph compiled from an architectural house is presented in Figure~\ref{fig:fig2}.

\subsection{CBIM cloud server}
\subsubsection{Cloud connection}
The technicalities of the CBIM system are designed to be transparent to general CBIM users. CBIM functions are embedded into the cloud server while the operations are reflected in local BIM software, providing users with CBIM coordination functions from within their software panel. Ideally, users would barely experience the existence of CBIM but enjoy the advanced functionalities it enables like geometry reference. In addition, the future CBIM can be licensed with a Software-as-a-Service (SaaS) approach by which users can access the service directly through a web browser to visualize models, communicate, collaborate, etc., without the need for installing and maintaining the application manually.
In this study, the CBIM server is implemented on a workstation external to the local BIM editors. The server combines a graph-based database and a series of semantic enrichment algorithms. A private network connection is established using \textit{Socket}, a python library, to link client computers and the CBIM server. The private network ensures information transmission security and is suitable for users within the same local network.

\subsubsection{Linking domain-specific graphs}\label{linkgeneration}
The CBIM server incorporates semantic enrichment algorithms to link, by adding meaningful relationships, standalone discipline graphs hosted on the database into an interconnected web of CBIM metagraph. As discussed in detail by \cite{ouyang2022semantic, wang2022cbim}, the rule-based algorithms utilize all information available from the CBIM data repository to infer and generate inter-domain relationship associations between building elements across federated sub-model graphs. 
The enrichment algorithms embed topology computation routines and rulesets that encapsulate expert knowledge on the underlying implications of building element placements. The entire process can be viewed as making explicit the implicit object relationship that an experienced engineer can infer upon observing the discipline models. Such semantics are believed to be the basis for various software methods to augment and automate design processes, while this study aims at demonstrating the use of the enriched \textit{CBIM:RelSpatial} relationships which register the computed topological properties between pairs of objects in facilitating asynchronous design collaboration across design disciplines.

\subsubsection{Geometry reference mechanism}
\begin{figure*}[ht]
	\centering
  	\includegraphics[width=\textwidth]{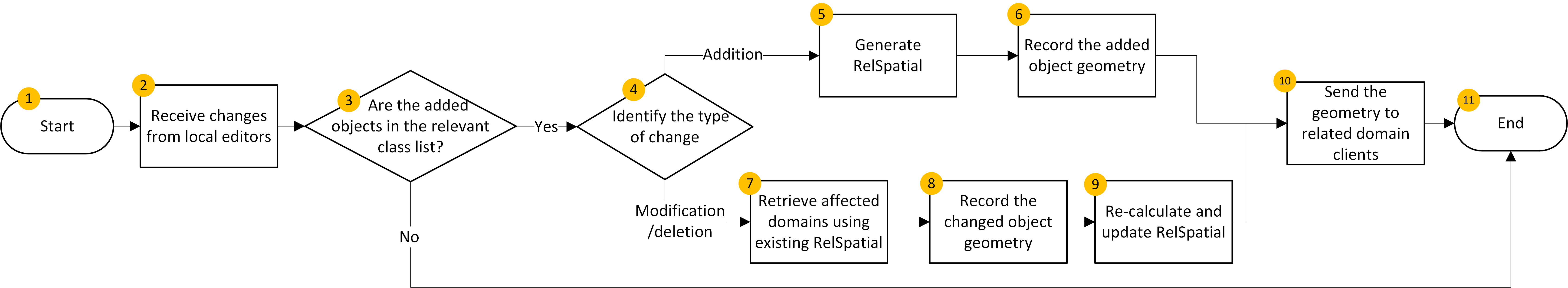}
  	\caption{Mechanism of preparing and sending object geometry for assisting asynchronous design.}
  	\label{fig:fig3}
\end{figure*}

When multidisciplinary graphs are linked appropriately by CBIM:RelSpatial relationships, one possible direction is to utilize them for assisting across-domain design tasks. 
For example, a common part of the design process is to exchange files for communicating design changes and updates using current design exchange technologies. 
However, if information regarding model updates from one domain can be sent to other related domain clients through a cloud server and that modified objects can be automatically rendered in other participants' interfaces, much time and human resources can be saved in communicating files and identifying changes. 
We believe that finding a mechanism to identify the relevance among objects across-domain is a key for the usability of the envisioned function, since then the system would be able to distinguish changes that will have a subsequent impact on other disciplinary design from those that do not, and propagate only the necessary ones to avoid unwanted disturbance. 

However, determining the relevance between objects from different domains is a complex task. If an architect decides to relocate an architectural wall, the columns hosted by the wall and the pipes attached to the wall, if exist, are potentially under impact. The responsible engineers should therefore be notified. 
Similarly, when an architect decides to extend the length of a wall, engineers need to consider adding extra columns or relocating existing ones for more balanced loading. 
Geometry-related modifications are a common category for reference propagation, but it also depends on object types. For example, furniture objects should be excluded from the propagation as structural engineers will likely find their changes irrelevant and prefer not to receive notifications and geometry references regarding these objects. 
Additionally, architect's adjustment on the type of a space function should also be communicated. For instance, if a room that was once labelled as a office is now changed to a library, the structural design would need to be re-evaluated to accommodate for the increased loading conditions. 
The examples listed above illustrates a few factors affecting the relevance of across-domain objects, such as object types, attribute types (geometry attributes or other semantic attributes), and disciplines (architecture, structure, MEP).


In this study, we are seeking to explore the feasibility of using the generated CBIM:RelSpatial relationships to determine relevance and achieve automatic geometry reference. We narrow down the scope to the architecture and structure domains, and aim to address location and dimension changes. The geometry reference mechanism is presented in Figure~\ref{fig:fig3}.

When the CBIM server receives changes from local software (2), a decision module evaluates whether the added objects are relevant for processing (3). For example, architectural objects that have no bearing on the structure, such as furniture, are excluded. A second decision module then identifies the type of change (4). 
If the change concerns addition of new BIM objects, the CBIM server executes semantic enrichment algorithms to correctly classify the new object and to associate it with existing objects, if applicable, by instantiating appropriate spatial relationships (5).  And the added object geometry is recorded and sent to related clients (6, 10).
For changes concerning modifications and deletions of existing objects, CBIM will first retrieve the affected domains by using the existing RelSpatial (7) and record the modified object geometries (8), followed by sending the collected geometry references to the discipline clients who are impacted (10). CBIM will also re-calculate the existing spatial relationships (9) to ensure that the CBIM database is up-to-date with the latest building designs. 
The procedure leverages the advantages of graphs and utilizes explicit inter-domain association relationships to determine domains and objects affected by a change action. 
Upon completing the procedure, the cloud platform will switch to a stand-by mode and await for the next cycle of change detection.
 
\section{Case Study}
\begin{figure}[H]
	\centering
  	\includegraphics[width=1.0\textwidth]{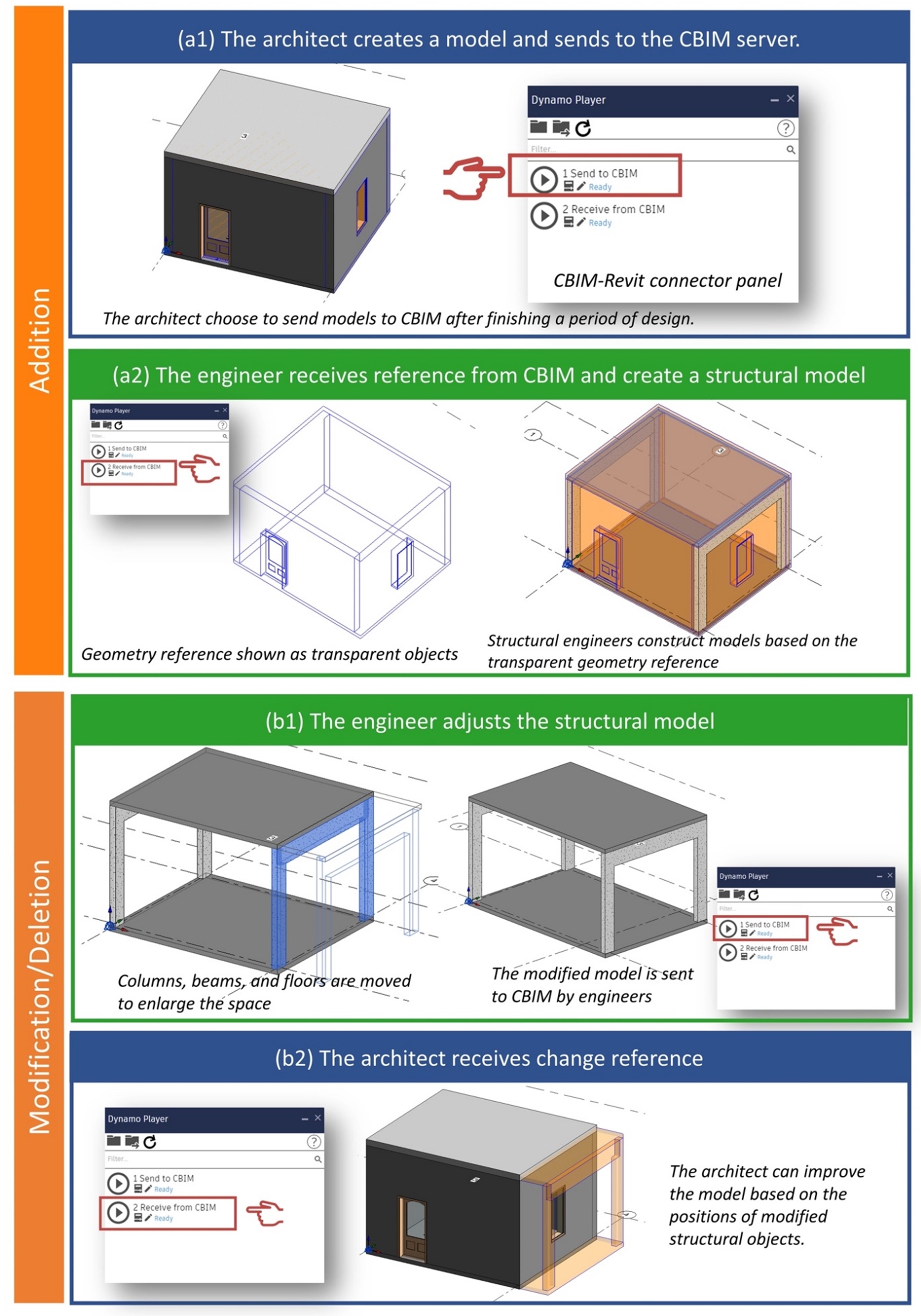}
  	\caption{Examples of using the CBIM platform for assisting across-domain asynchronous design.}
  	\label{fig:fig4}
\end{figure}

A typical federated collaboration in the design phase of a construction project usually begins with the architects, who design architectural models and deliver them to structural engineers as references. As long as change happens, participants need to export the latest model from their native design tools as neutral files and send them to corresponding design teams to ensure model consistency being maintained among all disciplines. A federated apartment project is selected to illustrate the fileless collaboration functionality enabled by the CBIM platform, as shown in Figure~\ref{fig:fig4}. In this case study, the architectural Revit and the structural Revit are installed on two separate workstations. The CBIM server is implemented on a third computer that connects to the two clients within a local network. 

The first design scenario is the addition of new objects. After the architects finish their architectural design in the architectural Revit, they upload their model data to the CBIM server using the CBIM-Revit connector (Figure~\ref{fig:fig4}~(a1)). Upon receiving the data, the geometry reference mechanism within the server is triggered to instantiate inter-domain spatial relationships for the new objects through running semantic enrichment algorithms. Since all objects are newly created and are in dependency with the structural design, their geometries are packaged by the module and sent to the structural editor. In other words, CBIM understands that the created architectural elements are essential in guiding the structural designs. The CBIM plugin within the structural Revit renders the received geometries to provide precise references for designing the structures (Figure~\ref{fig:fig4}~(a2)).

Another scenario concerns the modification and deletion of existing objects. As per the owner’s request, the engineer decides to relocate the columns and beams on the eastern façade to enlarge the interior space of the house. The engineer then pushes the updated design onto the CBIM server (Figure~\ref{fig:fig4}~(b1)). Again, the CBIM server activates the geometry reference mechanism and retrieves geometries of the structural objects in spatial associations with objects from the architectural model. These objects are considered as having a consequential impact on the architectural design and therefore need to be propagated. Geometry reference is sent to the architectural editor for rendering, based on which the architects can correct their design models manually (Figure~\ref{fig:fig4}~(b2)).

The across-domain communication process is fileless. Benefiting from the CBIM connector inside Revit, modeling information can be retrieved and sent to the CBIM server with a click of a button. In addition, the local network connection ensures that data is seamlessly transmitted from one platform to another. Furthermore, unlike the previous practice of packaging all model information as files, the CBIM approach compares design versions, extracts and delivers only the changed objects, realizing object-level design coordination.

From a technology aspect, the proposed pipeline can be implemented in a synchronous manner in the sense that CBIM is capable of monitoring and propagating changes in real time without user intervention. However, the culture of AEC design collaboration and the inherent industry practice may prefer an asynchronous route, since users need their independencies in terms of time and resource planning to participate in federated collaboration \cite{esser2022graph}. In this case study, an asynchronous approach for CBIM collaboration is implemented where changes are communicated only when users decide to push or pull to or from the server by clicking the button.

\section{Discussion}
 
This study implements a series of components to construct the CBIM platform and constructs a geometry reference mechanism to achieve object-level design collaboration across AEC disciplines.
Participants can benefit from the fileless asynchronous function that reduces human effort in exchanging design files. 
The automatic geometry reference mechanism sends the latest model modifications to relevant clients for supporting their design and help users get rid of manually identifying changes. Another advantage comes from the CBIM-Revit graph connector. It provides another approach to representing BIM models as graphs by retrieving raw data from a commercial BIM application directly. 
The connector is open for research use and would benefit researchers with the same interests. 

This study also faces a number of limitations. 
The connector is restricted to the Revit software environment, though the development of add-ins for other BIM software can follow the same principles and is purely technical. 
Secondly, the current implementation uses a private network, while a system that can be deployed on the public network would be preferable to facilitate wider access. 
Lastly, the current geometry reference mechanism might still send irrelevant notifications to clients since the current relevance determination process is primarily based on object topologies. The mechanism is also restricted to processing location-related and dimension-related changes. Others, like space type changes, cannot be dealt with yet.

The mentioned limitations lead us to our vision of a improved, more intelligent approach in which object relevance can be explicitly represented as constraints among BIM objects in a graph format. To achieve that, the first thing is to construct a constraint ontology that encapsulates different types of relevance relationship of BIM objects across disciplines on the class level. 
 The construction of the ontology can be done through interviewing experts and summarize their knowledge regarding the constraining object behaviors as an ontology. 
It would also be possible to use unsupervised machine learning techniques to cluster pairs of object nodes from different domains that might have underlying constraint meanings. Compared with manual extraction, the machine learning approach would be automatic, dynamic and may reveal some constraints that are omitted by humans, but it requires a large amount of data and well-prepared algorithms. 
After developing the constraint ontology, the second question concerns the instantiation of the constraint relationships. Rule-based algorithms would be appropriate to generate relationships that can be explicitly represented by machine-readable conditions. Graph neural networks, on the other hand, have shown strong learning ability for space type classification in previous experiments \cite{wang2021room, wang2022exploring}, and therefore may perform well when instantiating abstract constraint links. 
Eventually, multidisciplinary graphs will be linked properly by various types of relationship instances. Another software module can be developed to leverage the existence of explicit constraints and implement them to achieve intelligent functions, like intelligent change propagation for resolving design conflicts. 
 

\section{Conclusions}
The study proposed a set of tools to implement the CBIM platform and devised a case study to demonstrate the fileless multidisciplinary collaboration CBIM enables. The implemented platform supports model coordination through a linked, object-level graph database to help users eliminate the cumbersome file exchanges common to nowadays design workflows. The new paradigm improves design collaboration efficiency and allows users to focus on their design works instead of wasting time in communicating changes and transmitting files.

As the primary contribution, this study validates the feasibility of enabling cross-domain asynchronous design using the linked graph representations of multidisciplinary BIM models – a concept under the umbrella of the broader CBIM paradigm. The second contribution is the proposed geometry reference mechanism which utilizes the CBIM:RelSpatial relationships and object types to determine the relevance of across-domain BIM elements to process changes and prepare appropriate geometry references for corresponding design teams. The last contribution concerns the graph connector that directly compiles native Revit models into their CBIM format following a graph-based CDE approach. The connector is available for public research use at \url{https://github.com/ZijianWang1995/CBIM-Revit_Graph_Compiler}.

\section*{Acknowledgments}
This work is part of the Cloud-based Building Information Modelling (CBIM) project, a European Training Network. The CBIM project receives funding from the European Union's Horizon 2020 research and innovation programme under the Marie Skłodowska-Curie grant with agreement No 860555.

\bibliographystyle{unsrt}  
\bibliography{references}

\end{document}